\documentclass[12pt,preprint]{aastex}
\usepackage[tbtags,fleqn]{amsmath}
\usepackage{amsfonts}

\slugcomment{Accepted for publication in the Astrophysical Journal}

\shorttitle{Schmidt's Conjecture and Star Formation}
\shortauthors{Lada, Lombardi, Roman-Zuniga, Forbrich \& Alves}

\def\msun{M$_\odot$}

\def\13co{$^{13}$CO}

\def\Hs{\mathrm H}
\def\diff{\mathrm d}
\def\psd{$\Sigma_*(A_{\mathrm K})$}
\def\sfrsd{$\Sigma_{\mathrm SFR}$}
\def\ak{$A_K$\ }
\def\gsd{$\Sigma_{\mathrm gas}\ $}
\def\sdgta{$\Sigma_*(>A_{\mathrm K})$}
\def\sagta{$S(>A_{\mathrm K})\ $}
\def\sagtans{$S(>A_{\mathrm K})$}
\def\gta{$(>A_{\mathrm K})\ $}
\def\gtans{$(>A_{\mathrm K})$}

\begin{document}


\title{Schmidt's Conjecture and Star Formation in Molecular Clouds}

\author{Charles J. Lada }
\affil{Harvard-Smithsonian Center for Astrophysics, 60 Garden Street
Cambridge, MA 02138, USA }
\email{clada@cfa.harvard.edu }

\author{Marco Lombardi}
\affil{University of Milan, Department of Physics, via Celoria 16, I-20133 Milan, Italy}
\email{marco.lombardi@unimi.it}

\author{Carlos Roman-Zuniga}
\affil{Instituto de Astronom\'ia, Unidad Acad\'emica en Ensenada, Universidad Nacional Aut\'onoma de M\'exico, Ensenada, Ensenada BC 22860, Mexico}
\email{croman@astrosen.unam.mx}

\author{Jan Forbrich }
\affil{Institute for Astronomy,  University of Vienna, T\"urkenschanzstrasse 17, 1180 Vienna, Austria}
\email{jan.forbrich@univie.ac.at}

\and

\author{Jo\~ao F. Alves}
\affil{Institute for Astronomy,  University of Vienna, T\"urkenschanzstrasse 17, 1180 Vienna, Austria}
\email{joao.alves@univie.ac.at \vskip 2.0in}

\begin{abstract}
We investigate Schmidt's conjecture (i.e., that the star formation rate scales in a power-law fashion with the gas density) for four well-studied local molecular clouds (GMCs). Using the Bayesian methodology we show that a {\it local} Schmidt scaling relation of the form $\Sigma_{*}({A}_{\mathrm K}) = \kappa {A}_{\mathrm K}^{\beta}$ (protostars pc$^{-2}$) exists within (but not between) GMCs. Further we find that the Schmidt scaling law, by itself, does not provide an adequate description of star formation activity in GMCs. Because the total number of protostars produced by a cloud is given by the product of $\Sigma_*{( A_{\mathrm K})}$ and $S'(> A_{\mathrm K})$, the differential surface area distribution function, integrated over the entire cloud, the cloud's structure  plays a fundamental role in setting the level of its star formation activity.  For clouds with similar functional forms of $\Sigma_*(A_{\mathrm K})$, observed differences in their total SFRs are primarily due to the differences in $S'(> A_{\mathrm K})$ between the clouds. The coupling of $\Sigma_*{( A_{\mathrm K})}$ with the measured $S'(> A_{\mathrm K})$  in these clouds also produces a steep jump in the SFR and protostellar production above $A_K \sim$ 0.8 magnitudes. Finally, we show that there is no {\it global} Schmidt law that relates the star formation rate and gas mass surface densities between GMCs. Consequently, the observed Kennicutt-Schmidt scaling relation for disk galaxies is likely an artifact of unresolved measurements of GMCs and not a result of any underlying physical law of star formation characterizing the molecular gas

\end{abstract}



\section{Introduction}

\noindent
The construction of the Milky Way and other galaxies from rarified, gaseous material into immense regular systems of hydrogen burning stars is a complex physical process which has operated over most of cosmic history and is not yet fully understood. Development of a predictive  theory of star formation is an essential key to  the piecing together a complete picture of galaxy formation and evolution. A  fundamental achievement of any such theory would be to obtain an understanding of the physical processes that control the rate of star formation in interstellar gas. An important step toward achieving such an understanding is to empirically establish the underlying relation that most directly connects the rate of star formation to some physical property of the interstellar gas. A little more than a half century ago Schmidt (1959) considered this problem and formulated the following conjecture: ``It would seem most probable that the rate of star formation depends on the gas density and...that the number formed per unit of time varies with a power law of the gas density". Here we infer that when referring to  density Schmidt meant {\it surface} density since he then proceeded to argue that such a power-law relation between the star formation rate and gas  surface densities applied to  the solar neighborhood and furthermore that the index of this power law was equal to 2. Over the last few decades considerable effort has been devoted to investigations of Schmidt's conjecture on galaxy wide scales. Power-law relations between the surface densities of the star formation rate and gas mass have been found to describe star formation across entire galaxies and galactic nuclei,  but typically these relations have been characterized by a  less steep power-law index ($\sim$ 1.4-1.6) than originally proposed by Schmidt  (e.g., Kennicutt 1988; Kennicutt \& Evans 2013, and references therein). However, since his original paper, very little work has been devoted to investigating Schmidt's conjecture on more local scales.  

Knowledge of the star formation process has improved considerably over the past fifty years. Millimeter-wave, molecular-line observations have long ago established molecular clouds as the primary sites of star formation in the Galaxy and  infrared observations from the ground and most recently from space have enabled systematic studies of Young Stellar Objects (YSOs), sources in the earliest stages of star formation and stellar evolution within molecular clouds. As a result, there has been a revival of interest in Schmidt's conjecture and, in particular, the question of whether it can provide a useful description of star formation in galactic molecular clouds (e.g., Heiderman et al. 2010; Gutermuth et al. 2011; Lombardi et al. 2013). In order to address this question, we investigate here the extent to which the Schmidt's conjecture of a power-law relation between the {\it surface} densities of the star formation rate, $\Sigma_{SFR}$, and the gas, $\Sigma_{gas}$ can describe the star formation activity within individual molecular clouds.
We will refer to this relation as the internal Schmidt law when measured within individual molecular clouds (and the global or Kennicutt-Schmidt law when determined for entire galaxies). In practice, the quantities that are directly measured in galactic  molecular clouds are the surface density of YSOs and the dust extinction\footnote{ These quantities can be readily converted into $\Sigma_{SFR}$ and $\Sigma_{gas}$ with knowledge of the typical protostellar mass and age and the gas-to-dust ratio. Here
we adopt $\Sigma_{SFR}$ = $10^{-6} \Sigma_*$  (\msun yr$^{-1}$pc$^{-2}$) and \gsd $=$ 197$A_K$ (\msun pc$^{-2}$.)}. Recently, Lombardi et al. (2013) developed a Bayesian method for fitting parametric density models to discrete observations and applied it to evaluate the internal Schmidt law for the protostellar population in the Orion molecular cloud.   Lombardi et al. (2013) validated their methodology using numerical simulations of synthetic protostellar populations and demonstrated that their technique returned accurate measurements of the input parameters prescribed for the simulated protostellar populations.  Applied to the observations of the Orion A cloud 
their methodology resulted in relatively robust measurements of the star formation scaling law and its parameters in that cloud. Specifically they derived: $\Sigma_* = 1.65 (\pm 0.19) A_K^{2.03 (\pm 0.15)}$ (protostars pc$^{-2}$). Moreover, they found no evidence for a discrete extinction threshold in the internal Schmidt relation nor evidence for any measurable diffusion of the protostars from their immediate birthplaces.

In this paper we extend the analysis of Lombardi et al. (2013) to investigate three additional local clouds (i.e., Orion B, Taurus, and California) in order to determine if a Schmidt scaling relation can describe the star formation in these clouds and, if so, to then derive the parameters of that relation for direct comparison with those for Orion A. We will show that all three clouds are characterized by  Schmidt relations with well determined parameters. In particular, for two clouds the power-law index and coefficient of the relation are essentially the same as those of Orion A. However, in contrast to Orion A, there are two clouds in our sample for which the local Schmidt relation appears to be characterized by discrete threshold extinctions for the star formation surface density. In addition we will argue that the Schmidt scaling law, by itself, is not sufficient to provide a complete description of star formation in a cloud. We will show that the level of star formation within a cloud is instead given by the product of the local Schmidt law and the cloud's differential area distribution function, integrated over the entire cloud. Therefore, detailed knowledge of a cloud's structure is  critical to obtaining a complete description of its star formation activity. Moreover, because variations in cloud structure, particularly at high extinctions, are significant, the appropriate scaling law to describe star formation between clouds is the relation between the spatially integrated star formation rates and masses of the clouds (e.g., Lada et al. 2010, 2012). Indeed, we will further demonstrate that there is no Schmidt scaling law between molecular clouds due to the well-known scaling between mass and radius of such clouds. Finally we will discuss our results in the context of the extragalactic or global Schmidt law. We will argue that the observed  Kennicutt -Schmidt relation for nearby galaxies is an artifact of unresolved measurements of molecular clouds and not due to any underlying law of star formation.

\section{The Schmidt Scaling Law in Local GMCs}

\subsection{Sample and Data}

We selected four objects from the local cloud sample of Lada et al. (2010) for analysis: the Orion A, Orion B, Taurus and California molecular clouds. These four clouds were selected because they had the most complete published positional information and source classifications of all the clouds studied by Lada et al. (2010).  Source catalogs from Megeath et al. (2012), Rebull et al. (2011) and Harvey et al. (2013) were used for the Orion, Taurus and California clouds, respectively.  Infrared extinction measurements were taken from extinction maps derived from the 2MASS  sky survey using the NICEST algorithm (Lombardi et al. 2010a, 2011). The extinction maps were masked so that the areas considered corresponded to the spatial boundaries of the infrared surveys we used to identify the protostellar populations of the clouds. In all three clouds only sources classified as protostars, that is, Class I or Class 0 objects, were examined. This insured a sample of young objects likely still at or close to the locations of their formation. The Orion A, Orion B and Taurus clouds are characterized by relatively high star formation rates and numbers of protostars  and good statistics, while the California cloud, a GMC comparable in mass to Orion A, is characterized by a relatively low star formation rate and less robust statistics (Lada et al. 2009). Comparison of the Orion A and California clouds was considered to be particularly useful to examine the possible effect of a varying Schmidt-like scaling law in determining the very different total star formation rates in these otherwise similar clouds.

\subsection{Bayesian Analysis}

We used the Bayesian method developed by Lombardi et al. (2013) to investigate the relation between the {\it protostellar} surface density distribution, $\Sigma_*$, and the dust surface density distribution
measured by A$_K$, the infrared extinction, in the three clouds in our sample. Using Bayes' theorem we address the following problem: given a set of protostellar positions $\{ x_n\}$, the corresponding  extinctions, $\{(A_K)_n\}$ at those positions and a model for the protostellar surface density, $\Sigma_*(x | \theta)$, what can we infer about any parameters $\theta$?  Following Lombardi et al. (2013) we begin with a Schmidt-like model for $\Sigma_*(x| \beta, \kappa)$
of the form:

\begin{equation}
  \label{eqn1}
\Sigma_*(A_{\mathrm K}) = \kappa {\rm A}_K^\beta(x)
\end{equation}
\noindent
where $\Sigma_*(A_K) \diff S$ is the number of protostars in the area $\diff S$, which is characterized by
an extinction $A_K$.
Additionally we modify the model to allow for the possibility of a star-formation surface density
threshold, i.e., a lower limit for the extinction below which little or no star formation takes place and few if any protostars are produced in situ (e.g., \citealp{2010ApJ...724..687L},
\citealp{2010ApJ...723.1019H}). Finally we will also allow for some diffusion of protostars from their
birth sites.  We model this diffusion process by smoothing the initial
protostellar surface density, $\Sigma_*^{(0)}$, by a
Gaussian spatial kernel. In summary:

\begin{equation}
  \label{eq:2}
  \Sigma_*(x) = \int \frac{1}{2 \pi \sigma^2} \ e^{|x -
    x'|^2 / 2 \sigma^2} \Sigma_*^\mathrm{(0)}(x') \,
  \diff^2 x' \; ,
\end{equation}

where:
\begin{equation}
  \label{eq:3}
  \Sigma_*^\mathrm{(0)} (x) = \kappa \Hs(A_K(x) - A_0
  \bigr) \left( \frac{A_K(x)}{1 \mbox{ mag}} \right)^\beta \; .
\end{equation}

In this equation $\Hs$ is the Heaviside function
\begin{equation}
  \label{eq:4}
  \Hs(z) =
  \begin{cases}
    1 & \text{if $z > 0 \; ,$} \\
    0 & \text{if $z \le 0 \; .$}
  \end{cases}
\end{equation}

\noindent
 Here $\kappa$ is the normalization constant, or star formation coefficient (measured in units of $\mbox{star pc}^{-2} \mbox{mag}^{-\beta}$), $A_0$ is the star formation threshold (in units of magnitudes of $K$-band extinction), $\beta$ is the dimensionless exponent, and $\sigma$ is the diffusion coefficient (measured in pc).

Starting with an assumed {\it prior} distribution of parameters, 
 $\theta  = \{\beta$, \  $\kappa$, $\sigma$, A$_0$\},
 we can use Bayes' theorem to derive the {\it posterior} probability distribution $P(\theta | \{x_n\})$ and the desired distribution of parameters, once the observations $\{x_n\}$ have been made.  Lombardi et al. (2013) showed that the likelihood appearing in Bayes'  theorem, 

\begin{equation}
  \label{eq:5}
  P\bigl(\theta | \{ x_n \} \bigr) = \frac{\mathcal{L}\bigl( \{ x_n \}
    | \theta\bigr) p(\theta)}{\int \mathcal{L}\bigl( \{ x_n \} |
    \theta' \bigr) p(\theta') \, \diff \theta'} \; \cdot
\end{equation}

\noindent
can be written as:
\begin{equation}
  \label{eq:6}
  \ln \mathcal{L}\bigl(\{x_n\} | \theta\bigr) = \sum_{n=1}^N \ln
  \Sigma_*(x_n | \theta) - \int \Sigma_*(x | \theta) \, \diff^2 x \; .
\end{equation}

\noindent
Using Equation~(6) we inferred the four parameters $\theta$ using flat priors over all of them.  The posterior probability distribution was explored with a Monte Carlo Markov Chain integration using a simple Metropolis-Hastings sampler. To implement the Bayesian analysis using our data we determined the extinction in the pixel containing each protostar and associated that extinction with that protostar. The method is not affected by the existence of more than one protostar in a given extinction pixel. We consider only non-masked pixels
but both those with and without protostars.   The results  are presented in the next section below. 

\subsection{Results: The Star Formation Law in Local Clouds}

Table 1 lists the sixteen posterior parameters for the model, Equation (2), derived for the four clouds using our Bayesian analysis. The values for Orion were previously published in Lombardi et al. (2013). Figures 1, 2, 3, and 4 show surface density plots of the posterior probabilities for all combinations of the four parameters, $\beta$,  $\kappa$, $\sigma$ and $A_0$ in Orion A, Taurus, California and Orion B, respectively.  The data in the table and figures provide compelling evidence that a Schmidt-type scaling relation can describe star formation within these clouds. The values of the parameters in Equation (2) are well constrained in 15 of the 16 calculated posterior probabilities (the derived probability distribution for $A_0$ in Orion B being the exception).  

The small values derived for  the diffusion coefficient, $\sigma$, suggest that the protostars have not drifted very far from their birth places over their lifetimes. This is consistent with the identification of these sources as extremely young objects and with the idea that they may still be protostars accreting material from their surroundings. An upper limit for the drift distances is set by the physical scale of the pixels in the Nyquist sampled extinction map for each cloud. This scale ranges from 0.05 pc in Taurus to 0.17 pc in the Orion clouds. The corresponding range in upper limits  of the protostellar drift velocities are $v_{drift} < 0.2 - 0.7 \ {\rm km\ s^{-1}}$ assuming a protostellar age of 0.25 Myr. Because the values for $\sigma$ in all four clouds are essentially equal to zero we will ignore this parameter in the expression for the star formation scaling law from here forward. 

In Equation (1) the star formation coefficient, $\kappa$, sets the overall scale of star formation in the molecular gas and $\beta$ governs how the level of star formation varies with column density. These two parameters must be set to some degree by the underlying physics of the star formation process itself. The values for $\beta$ and $\kappa$  are found to be in surprisingly close agreement in three (Orion A, Taurus, California) of the four clouds, suggesting a common nature for the parameters in these clouds.  Using weighted averages for $\beta$ and $\kappa$ we can now write Equation (1) for these three clouds as:
\begin{equation}
  \label{eqn7}
\Sigma_*(A_K) = 1.7 \times  A_{\mathrm K}^{2.0}\ \ \ \ \ \  {\rm stars\  pc^{-2}}
\end{equation}

\noindent
We can also express the above relation in a form more similar to the standard Schmidt Law, that is, in terms of the star formation rate surface density, $\Sigma_{SFR}$, and the total (H $+$ H$_2$ $+$ He $+$ ...)  gas surface density, $\Sigma_{gas}$:

\begin{equation}
 \label{eqn8}
\Sigma_{SFR} = 4.6 \times 10^{-11} \times \Sigma_{gas}^{2.0} \ \ \ \ \ \ {\rm M}_\odot\  {\rm yr}^{-1}\ {\rm pc}^{-2}\
\end{equation}

\noindent
using the extinction law of Rieke and Lebofsky (1985), a normal gas-to-dust ratio (i.e.,  $N(H) = 2 \times 10^{21} A_V$ cm$^{-2}$), and a mean mass per H particle of $\mu$ = 1.36, corresponding to a hydrogen abundance by mass of 73\% (Allen 1973) and additionally assuming a typical protostellar age and mass of 0.25 Myr  and 0.25 \msun , respectively. 

The parameters $\kappa$ and  $\beta$ differ for the Orion B cloud, and the resulting star formation law for this cloud is given by:
\begin{equation}
\label{eqn9}
 \Sigma_*(A_{\mathrm K}) = 0.77 \times A_{\mathrm K}^{3.3}\ \ \ \ \    {\rm stars\ pc^{-2}} 
\end{equation}
\noindent
It is interesting to note that the steeper dependence of $\Sigma_*$ on $A_{\mathrm K}$ in Orion B compared to the other three clouds  is somewhat compensated for by the significantly smaller value of $\kappa$. Thus, $\Sigma_*(A_{\mathrm K}$) in the Orion B cloud (i.e., Equation 9) only exceeds that in the Orion A, Taurus and California clouds (i.e., Equation 7) at extinction levels of $A_{\mathrm K} >$ ~2.0 magnitudes. In the Orion B cloud only 1\% of the cloud mass is found at such high extinctions. Thus despite the steeper dependence of $\Sigma_*$ on extinction, the Orion B cloud is actually less effective in producing protostars at extinctions below 2.0 magnitudes than the other clouds in the sample.

We do not explicitly include the threshold parameter, $A_0$, in Equations 7, 8 or 9. Only two of our clouds, the California Molecular Cloud  and the Orion B cloud, showed any evidence for a sharp threshold extinction for $\Sigma_*$. The threshold for the California cloud is detected at high confidence while the detection of a threshold for the Orion B cloud is a more marginal result. In this latter cloud, the greater uncertainty in this parameter is largely due to the fact that the Spitzer protostellar survey of Orion B (Megeath et al. 2012) is the spatially least complete of the clouds studied here and covers only a relatively small portion of the low extinction region of the cloud. 
 Both the Orion A and Taurus  clouds, the two best studied objects here, showed no indication of a such a sharp threshold in the derived posterior values of $A_0$.  As discussed in Lombardi et al. (2013), detection of an extinction threshold depends on secure identifications of protostars particularly in regions of low extinction where the numbers of protostars are low due to the non-linear dependence of $\Sigma_*$  on extinction. It is possible that in these latter regions contamination from mis-identified Class II YSOs and background galaxies could mask the presence of such a threshold. However a more detailed assessment of the natures of the sources identified as low extinction protostars in Orion and Taurus would be needed before the possible existence of a discrete extinction threshold in those sources could be more seriously considered. 

Although the finding of a general Schmidt-like scaling relation for molecular clouds in this and earlier studies (i.e., Heiderman et al. 2010; Gutermuth et al. 2011) potentially provides significant insight into the process of star formation and may be useful as a predictive tool for studies of star formation in other contexts, it is important to realize that this relation, by itself, does not provide a complete description of the overall level of star formation characterizing the clouds. This is illustrated in Figure 5. The left hand panel shows the Schmidt relation ($\Sigma_*$ vs $A_K$) for the Orion A cloud.  Here the observed $\Sigma_*(A_K)$  steeply rises in an unabated fashion to the highest measured extinctions in the cloud. Also plotted is the least-squares fit to the data whose derived parameters of $\beta = 2.0\pm 0.13$ and $\kappa =1.4 \pm 0.14$ stars pc$^{-2}$ are essentially identical to those inferred  from our Bayesian analysis. In the right panel we plot the fraction, $N_*(>A_K) / N_*(total)$, of protostars observed above a given extinction, $A_K$.   Despite the fact that $\Sigma_*(A_K)$ is so steeply rising with extinction, the actual number of protostars produced by the cloud falls off sharply with extinction for values of $A_K > 0.8-1.0$ magnitudes. 

This seemingly paradoxical situation is a result of the facts that 1) the number of protostars produced at any extinction is the product of the protostellar surface density and cloud area at that extinction and 2) that molecular clouds are stratified with well-behaved surface density profiles that fall steeply with radius (e.g., Lada et al. 1999; Alves et al. 1998, 1999; Alves et al. 2001; Lombardi et al. 2010b; Arzoumanian et al. 2011).  In the next section of the paper we will examine the significant effect of this aspect of cloud structure on the overall level of star formation in a cloud and derive a more complete description for the star formation process in molecular clouds.

\section{The Total Star Formation Rate and the Crucial Role of Cloud Structure}

\subsection{The Integrated Star Formation Scaling Relation}

By itself, the Schmidt scaling relation does not appear to be a reliable predictor of star formation activity in local GMCs. Moreover, observations indicate that clouds of similar size, mass and average $\Sigma_{gas}$ can have total or integrated star  formation rates (SFRs) and global values of $\Sigma_{SFR}$ that vary by as much as an order of magnitude (Lada et al. 2010, see also Figure 8). Given the generally similar natures of the star formation laws in our cloud sample how is it possible to explain this variation?  To answer this question and address the issue of how a steeply rising star formation law produces a steeply declining population of protostars and SFR at large column densities, we have to explicitly take into account the relation between cloud column density and is surface area. The  number of protostars at a given level of extinction, $A_\mathrm{K}$, is the product of the area $S(A_\mathrm{K})$ encompassing that extinction and $\Sigma_*(A_\mathrm{K}$). The total number of protostars is given by  the integral of this product over all extinctions in the cloud.  

Suppose we know the \textrm{integral\/} relation between the projected surface
area of the cloud and the column density, expressed in terms of the
area of the cloud above a given extinction $A_K$.  Let us call this relation
the surface area distribution function, $S(> A_K)$.  Then the total number of protostars 
in the cloud is:
\begin{equation}
  \label{eq:9}
  N_*= \int \Sigma_*(A_\mathrm{K}) \, \diff S = 
  \int \Sigma_*(A_\mathrm{K}) \bigl| S'(>A_K) \bigr| \, \diff
  A_\mathrm{K} \; .
\end{equation}
This equation tells us that we can estimate the expected number of
protostars that a cloud will produce from the integral of the product of the
density of protostars as a function of $A_K$ and the \textit{differential\/} cloud
area. We  now propose that variations in the star formation rates {\it between} 
clouds are largely due to variations in the function $S(> A_K)$ and its derivative.

In Figure 6 we show plots of $S(>A_K)$ vs $A_K$ for the four clouds in this study and the Pipe 
Molecular Cloud for comparison. The figure shows that for all clouds $S(>A_K)$ is a decreasing function of $A_K$  and steeply falls at the higher extinctions. Large differences 
in amplitudes and shapes are apparent for the four sources, even on this log-log plot. These differences qualitatively appear to be correlated with the differing levels of star formation in the clouds, from Orion A, the most active, to the Pipe the least active star forming cloud. This confirms our intuition regarding the importance of $S(>A_K)$ in determining the level of star formation, given the similar nature of the local Schmidt law for these clouds. To further test this idea we evaluated the integral Equation (10) for each cloud with $\Sigma_*$ given by Equation (7) and the $A_K^\beta$s calculated from our extinction maps. The results are shown in Table 2 and the predictions agree very well with the observations. Our analysis also confirms our earlier suspicions (Lada et al. 2009, 2010) that differences in cloud structure, particularly at high extinctions, was the primary cause of the differences in total star formation rates between local clouds.

We note that in a recent paper  Burkert \& Hartmann (2013) analyzed data for a sample clouds in the Spitzer C2D survey reported by Heiderman et al. (2010) and  found the surface area of the clouds  to decrease rapidly with mass column density for the combined cloud sample, similar to what is found here.  They further suggested that this steep decline in cloud surface area drives the steep rise in \sfrsd \  in the Schmidt relation and moreover  posited that the variations in \sagta between clouds produce variations in the $\beta$s of the corresponding Schmidt relations. We find no evidence to support this suggestion in the local cloud data presented here. Instead, as pointed out earlier, we find similar Schmidt relations for clouds that display clear differences in \sagtans.

\subsection{The Concept of an Extinction Threshold for Star Formation}

We can gain more insight into the key role of \sagta by numerically evaluating the integrand of Equation 10 using a set of semi-empirical models with different values of $\beta$ and an assumed \sdgta. In Figure 7 we plot a series of model curves that represent the expected behavior of  $N_*$\gta with extinction for a set of discrete values of $\beta$ that range from 0 to 10. We have also assumed the observed \sagta for the Orion cloud (see figure 6) and normalized the profiles (i.e., $N_*$\gta / $N_*(total)$ ) to calculate the cumulative protostellar fraction (hereafter CPF)  and remove any dependence on $\kappa$.  Finally, for comparison we have plotted the data for Orion A. The predicted relations show how the CPF  will vary with increasing \ak for different Schmidt Laws, given a cloud with identical structure as Orion A. We now examine more closely some of the more informative aspects of these models for understanding the basic scaling relations for star formation in molecular clouds.
  
 At one extreme, consider the case of $\beta = 0$, which corresponds to a constant protostellar or SFR surface density, in other words, a cloud with no  Schmidt law. In this case the shape of the CPF vs \ak relation is predicted to be identical to that of the assumed normalized  $S(>A_K)$ profile (see Figure 6). The relation falls non-linearly with extinction until it reaches an extinction ($\approx$ 5.0 magnitudes), where it appears to be truncated. This truncation is a result of the fact that cloud column densities above this value are so rare that they are not detectable in a NICEST, 2MASS map of Orion A. This is, to some extent, a function of the spatial resolution of the observations and we would expect this steep cliff to move to somewhat higher values of extinction with observations characterized by better angular resolution. But at some point the fractional cloud area that exists at extreme extinction must be so low that the probability of finding a protostar there becomes vanishingly small. Indeed, for Orion with roughly 330 protostars, we would expect to find very few if any objects above the infrared extinction (3.0 magnitudes) where the fractional area of the cloud is about 0.3\% of the total or less, even though the total cloud mass measured above this level exceeds 800 \msun. 

We now examine the other extreme, where $\beta$ takes on a very large value. This is illustrated by the  model corresponding to $\beta = 10$. Here the predicted relation is characterized by a constant value of essentially unity (corresponding to $N_*$\gta $=$ $N_*(total)$)  for  all extinctions out to the truncation extinction of approximately 5.0 magnitudes where it  precipitously declines. In this situation the protostars appear to form only above a relatively sharp threshold extinction that marks the narrow column density  range of the remaining cloud area containing the highest average dust and gas column densities. Within this area star formation would be characterized by extreme protostellar and SFR surface densities, the quintessential conditions for cluster formation. 

Between these two extremes, that is,  for $\beta$s between 1 -- 4, the predicted curves are more or less similar in overall shape to the $\beta = 0$ case. This general resemblance attests to the critical importance of cloud structure in ultimately determining the total SFR even for clouds with such relatively steep power-law relations.  It is also apparent that as $\beta$ increases, the curves flatten and begin to approach the functional form of a $\beta >> 1$ curve, essentially the limiting case of a sharp or Heaviside-like extinction threshold in the protostellar production (or equivalently the integrated SFR). Below this threshold the influence of \sagta on the SFR is diminished. For example, the $\beta = 4$ model is essentially flat out to about 1.0 magnitudes before beginning an accelerated decline due to the increasing influence of the steeply falling  \sagta function.  Almost all the protostars in the model cloud appear to be forming above this threshold.  Even for the case of $\beta$ $=$ 2 the modeling predicts a relatively steep threshold, though at a lower extinction of around 0.5 magnitudes.

Finally, we are now in a better position to understand the earlier findings of Lada et al. (2010)
who found that the total star formation rate in local molecular clouds appeared well correlated with the mass of cloud material at high extinctions and suggested that this might indicate the existence of an extinction threshold for star formation near an \ak of 0.8 magnitudes. Recently, in separate studies, Ybarra et al. (2012) and Evans et al. (2013) reported very similar results for the Rosette Molecular Cloud and the Spitzer C2D  $+$ Gould's Belt sample of twenty-nine nearby dark clouds, respectively.  Lada et al (2010)  also used the steep Heaviside function to illustrate the possible idealized form of an extinction threshold for the spatially integrated star formation rate but cautioned that their data suggested a broad (factor of two) range in the threshold centered around an infrared extinction of 0.8 magnitudes suggesting, however, that such an ideal form might be difficult to achieve in nature. Comparison of the models (Figure 7) to the observations of the Orion A cloud  appears to provide strong general support of their hypothesis that the bulk of star formation is confined to the high (\ak $\ga$ 0.8 magnitudes,  \gsd  $\ga$ 160 \msun yr$^{-1}$) column density regions of GMCs. 

For example, in the $\Sigma_*$ -- \ak observational plane of the Schmidt law, as indicated by both our  least-squares  and Bayesian analyses, the Orion A cloud is characterized by a simple power-law with $\beta = 2$ and $A_0$ $=$ 0. Indeed the observed points in Figure 7 lie very close to the $\beta = 2$ curve in the CPF--\ak observational plane. However, inspection of  Figures 5 and 7 show that approximately 80\% of all the protostars in Orion are found above an $A_K$ of 0.8 magnitudes and 90\% above 0.5 magnitudes, even though lower extinction regions account for 90--95\% of the area of the cloud. Expressed in another way, the mean surface density of protostars in cloudy material  with dust column densities above 0.8 magnitudes is more than two orders of magnitude higher than that characterizing the cloud material at all lower extinctions. This jump in protostellar production, $N_*$\gtans,  in the vicinity of the 0.8 magnitude extinction boundary closely resembles a physical threshold and is reflected in the initiation of an increasingly steep downturn in the observed and predicted CPF relations between 0.5 and 0.8 magnitudes. 
The decline in the CPF  with extinction for Orion is not particularly sharp indicating again that in nature infinitely steep thresholds are difficult to produce. This is because, even if initially formed above such a threshold, a sufficiently small fraction of a cloud's protostellar population can: 1) migrate away from their birth sites, 2) displace surrounding material through winds and outflows, 3) be misidentified, or 4) form in rare cores, located in the outer regions of the cloud. In any event, though a threshold is clearly present, its precise location is difficult to quantify without some prior definition of an appropriate value for the CPF above which one considers a significant fraction of stars to be forming.  The fact that star formation activity is negligible at and above the truncation extinction of 3.0 magnitudes (due to the lack of cloud material at those high column densities) means that the bulk (75\%) of the star formation in Orion A takes place in a limited range of column density, between roughly 0.8 and 3.0 magnitudes of infrared extinction. The factors considered above effectively combine to produce a physically meaningful, albeit somewhat smooth, extinction threshold for star formation with clear measurable consequences (Lada et al 2010; Evans et al. 2013)\footnote{The measurement of a meaningful threshold for star formation requires the condition that the gaseous mass, $M_{cloud}$, contained in the measured cloud area, $S_{cloud}$, is larger than that,  $M_* = \sum_{n=1}^N(m_*(n))$,  of the summed masses, $m_*(n)$, of all the inner protostellar envelopes, themselves. These masses will vary from cloud to cloud but in the clouds studied here we find that the condition is satisfied at or near  the highest measured extinctions ($A_K \sim$ 3-5 magnitudes) in the individual clouds, well above the inferred SFR threshold of 0.8 magnitudes).}.

It is important to emphasize here that the concept of a threshold describing the integrated star formation activity  in the CPF -- \ak observational plane (Equation 10), is not equivalent to that of a discrete threshold in the $\Sigma_*$ -- \ak plane of the Schmidt relation (i.e, $A_0$ of Equation 3). This should be clear from the fact that in the Orion A cloud our analysis here and the earlier study of Gutermuth et al. (2011) find no measurable evidence for a threshold extinction in the areal Schmidt relation for Orion A.

\section{Comparison with Previous Studies}
\subsection{The Star Formation Law}

 Gutermuth et al. (2011) and Harvey et al. (2013) have previously explored the relation between the stellar and mass surface densites for the Orion and California clouds, respectively. Infrared observations obtained with NASA's Spitzer Space Telescope and ESA's Herschel satellite, respectively,  were used to compile nearly complete census' of young stellar objects (YSOs) in the clouds and infrared observations from 2MASS and Herschel were respectively used to measure the corresponding extinctions in the Orion and California clouds. Gutermuth et al. (2011) determined surface densities of YSOs from a nearest neighbor technique and included both protostars (Class I sources) and pre-main sequence stars (Class II sources) in their sample. They employed  a two-dimensional $\chi^2$ minimization technique to perform line fits to data in a log $\Sigma_{YSO}$ vs. log $\Sigma_{gas}$ plot and derived a value for $\beta$ of 1.8~$\pm$ 0.01 for combined observations of the Orion A and B clouds. If we similarly combine the Orion A and B data we derive a value of $\beta$  = 2.2 $\pm$ 0.07  from the Bayesian analysis of the  for protostellar sources, somewhat steeper than that of Gutermuth et al.  However, their derived value of $\kappa$ is significantly (approximately a factor of 25) larger than that we derive. This difference is interesting since, as discussed earlier, a least squares fit we performed to the $\Sigma_*$ vs $A_K$ relation using our data yielded a $\beta$ of 2.1 and a coefficient of $\kappa =$ 1.6 for the protostellar sources in Orion A, nearly identical to what we derive from the Bayesian analysis. We speculate here that the origin of this difference could result from two factors. First, we expect that inclusion of both Class I and II sources in the Gutermuth et al. fit would contribute to the derivation of a larger $\kappa$ since there are ten times as many Class II sources as protostellar Class I sources in the cloud. We can estimate the $\kappa$ that would result in this case with a mathematically robust  exercise similar to that used to generate Table 2 in the previous section. We evaluate the integral in Equation 10 by fixing $N_*$ to be equal to the total number of Class I and II sources in the Orion A and B clouds. We then set $\Sigma_* = \kappa A_K^{2.0}$ and solve for $\kappa$ by numerically integrating Equation 10.  We find $\kappa =$ 12, a value six times larger than that in Equation 7. Second, although we used essentially the same source catalog as Gutermuth et al., the methodologies of the two studies significantly differ, especially with regard to the extinction maps and this can lead to a difference in the derived coefficients. The angular resolution of the Gutermuth observations is defined by a (20th) nearest-neighbor distance for background stars and spatially varies with extinction. Their mean resolution (6.2 arc min) is a factor of two greater than that of the fixed value (3 arc min) for the NICEST Orion extinction map used in this paper. Moreover, the Gutermuth et al. resolution is further degraded relative to that used here in the regions of highest extinction where most of the protostellar sources are located.  The NICEST extinction maps used here can, for fixed resolution, probe significantly deeper extinctions and more effectively remove foreground star contamination than conventional extinction mapping techniques (Lombardi 2009).  This is reflected in the fact that the maximum extinction Gutermuth et al.  measure is $A_V$ $\approx$ 24 magnitudes compared to $A_V$ $\approx$ 45 magnitudes in the NICEST map used in this study.  The non-linear dependence of $\Sigma_*$ on the gas surface density will result in the measurement of a larger coefficient, $\kappa$, for the spatially degraded, conventional extinction maps and we estimate (from direct comparison of our NICER and NICEST maps) that this could increase the coefficient by a factor of 4-5 relative to the value derived from the NICEST map and perhaps account for the remaining part of the discrepancy in the derived values of $\kappa$.  Combined, these two effects can plausibly explain both the magnitude and direction of the difference in the derived coefficients and therefore we are not concerned by this difference between the two studies. 
  
 Harvey et al. used basically the same source catalog as used here for the California cloud, but smoothed their data to a 0.2$^o$ angular scale and plotted the ratio of ${\Sigma_*} \over {A_K}$ as a function of $A_K$ and found a steeply rising slope ($\approx$ 4) for the relation between the two smoothed quantities. The steep slope they derived is not confirmed by our Bayesian analysis. However, a least-squares fit to our data results in a value for $\beta$ of approximately 3. These differences are likely the result of the heavy spatial smoothing employed by Harvey et al. which acts to dilute the detectability of  the threshold column density found in our present study. Indeed, when we account for this threshold, a least squares fit to our data returns a value for $\beta$ $=$ 2, in agreement with the Bayesian analysis.
 %\footnote{For example, the standard line fitting techniques, such as used by Harvey et al. and Gutermuth et al. inherently assume that the entire range of observations can be fit by a single power-law function, thus only two parameters ($\beta$ and $\kappa$), can be returned from such fitting procedures, whilst the Bayesian technique enables simultaneous derivation of many parameters (e.g, $\beta, \kappa, \sigma, A_0$)} 
Indeed, these authors also remarked that at low extinctions ``...but still above the general  background level, essentially no YSOs are found'' (Harvey et al. 2013).
 
 Finally we note that Gutermuth et al. (2011) also derived power-law correlations between $\Sigma_{YSO}$ and $\Sigma_{gas}$ for seven additional individual clouds with power-law indices that ranged between 1.4 and 2.7.  They included both Class I and II sources in their fits and did not find as close an agreement in the derived parameters for the clouds within their sample as we have for the clouds studied here. Their sample consisted of clouds that are somewhat more distant than the ones studied here and as mentioned above their methodology differs from that in this paper. Nonetheless, their results are still consistent with the idea that Schmidt-like relations generally characterize the star formation within molecular clouds. 
 
 \subsection{Extinction Thresholds for Star Formation}

\subsubsection{The Internal Schmidt SFR Scaling Relation}

The first extensive study of the internal Schmidt scaling law within molecular clouds was that of  Heiderman et al. (2010). They combined direct observations of protostellar surface densities and infrared extinctions from a sample of 20 nearby cloud regions from the Spitzer C2D survey with far-infrared (FIR) luminosities and HCN molecular-line observations from a sample of more distant galactic clouds to construct  a merged plot of  $\Sigma_{\mathrm SFR}$ vs $\Sigma_{\mathrm gas}$ spanning both samples. They found a steep power-law rise of $\Sigma_{\mathrm SFR}$ with $\Sigma_{\mathrm gas}$ followed by a possible break or leveling off of the relation near an infrared extinction of about 0.65 magnitudes which they interpreted as indication of a threshold in the $\Sigma_{\mathrm SFR}$ at that extinction. The similarity of the inferred extinction thresholds for the  differential and integrated (Lada et al 2010) star formation rates suggested that they could have a similar origin, however at the time  the exact nature and relation of the two inferred star formation thresholds was not understood. 

Gutermuth et al. (2011) found no clear evidence for Heaviside-like thresholds or breaks in the $\Sigma_{\mathrm SFR}$ vs $\Sigma_{\mathrm gas}$ relations for 8 nearby clouds. Moreover, they  argued that without the addition of FIR/HCN data there would be no clear evidence for a break or threshold observed in the Heiderman et al. data.   Burkert \& Hartmann (2013) argued that the break observed in the Schmidt relation by Heiderman et al. was the result of combing observations of different cloud samples that have differing Schmidt relations due to physical differences in cloud structure (i.e., \sagtans).  Furthermore,  they suggested that a specific surface density threshold in the internal Schmidt law is not necessary to explain either the results of Lada et al. (2010) or Heiderman et al. (2010). In this paper we also examined the internal Schmidt scaling relation between the  protostellar surface density ($\Sigma_*$) and the dust surface density in four clouds and  found evidence for definite extinction thresholds in two of them, but for one of those the evidence was somewhat marginal.  So it is difficult to draw firm conclusions regarding how common the presence of such steep thresholds are in the typical Schmidt relations for molecular cloud given the state of existing studies. Analysis of a larger sample of clouds using Bayesian techniques would help to elucidate this particular issue. However, any effect of the presence or absence of such \psd \  thresholds on the overall production of protostars and the SFR  in local clouds appears to be insignificant compared to the effect of  \sagta on these basic properties of star formation within these clouds. In addition there is no evidence in our data to support the suggestion by Burkert \& Hartmann (2013)  that variations in cloud structure produce variations in the form of the internal Schmidt scaling laws.

\subsubsection{The Internal Integrated SFR Scaling Relation}

The results of this paper suggest that even if discrete thresholds in the Schmidt laws of molecular clouds were rare, a more ubiquitous type of threshold, one that characterizes the scaling law of integrated SFR vs dust column density, may offer a better description of star formation in local molecular clouds.  In particular, our analysis of the Orion A data demonstrates that the absence of a threshold, or break,  in the Schmidt relation has little bearing on the existence or fundamental significance of the thresholds in the CPF--\ak relation and also likely those inferred by the earlier Lada et al. (2010) and more recent Evans et al. (2013) studies. Consider that  in this study here we find no evidence for a Heaviside threshold in the Schmidt relation for the Orion A cloud, yet we found compelling evidence for a threshold-like behavior  in the integrated or total protostellar population (and SFR) in that cloud. For example, as briefly indicated earlier, the mean surface density of protostars above the 0.8 magnitude extinction boundary in Orion A is $<\Sigma_{*(A_K \geq 0.8)}>$ = 4.6 stars pc$^{-2}$ whilst below this boundary we find $<\Sigma_{*(A_K < 0.8)}>$ = 0.05 stars pc$^{-2}$, a difference approaching two orders of magnitude. Similarly, Evans, Heiderman \& Vutisalchavakul (2013) have recently shown that in the 29 clouds of the C2D and Gould's Belt Spitzer cloud sample, the mean $\Sigma_{\mathrm SFR}$ for Class I sources found at extinctions $A_V$ $\geq$ 8 magnitudes (i.e., $A_K \geq 0.9$ magnitudes) is 14 times larger that that measured for sources characterized by $A_V$ $<$ 8 magnitudes. Evans et al. also found the star formation rate above this threshold correlates linearly with the mass of gas above the threshold, similar to the original findings of Lada et al. (2010). We also note that millimeter and submillimeter continuum surveys of local clouds for dense cores, the present and future birthsites of protostars, have shown that such cores are rarely found in regions whose infrared extinctions are less than 0.8  magnitudes (e.g., Johnstone, DiFrancesco, \& Kirk 2004; Enoch et al. 2008; Andre et al. 2010). Indeed, Enoch et al. remark that ``There appears to be a strict extinction threshold in Serpens and Ophiuchus, with no cores found below  $A_{\mathrm V} \sim$ 7 and 15 mag., respectively.'' Combined, all these results confirm that there appears to be a physically meaningful threshold for star formation in local clouds. This threshold is not necessarily sharp and it is not due to, nor in the form of, a break in the internal Schmidt relation within the clouds. Instead, we propose here that it is in the form of a highly elevated (spatially integrated) star formation rate in gas above the threshold  likely resulting from the relatively steep non-linear rise of $\Sigma_*(A_{\mathrm K})$ with extinction coupled with the increasingly steep non-linear decrease and eventual truncation of $S'(>A_{\mathrm K})$ at high extinctions.

 \section{On the Global Schmidt Law in GMCs and Galaxies}

\subsection{Local GMCs}

In this section we consider the concept of a {\it global} Schmidt scaling relation for giant molecular clouds (GMCs).  We use the global or total  (spatially integrated) SFRs derived by Lada et al. (2010) for their local cloud sample. We then calculate $\Sigma_{SFR}$ for each cloud by dividing by the total cloud area (i.e., $S(>0.1 \rm{mag})$) determined from our extinction maps.  In Figure 8 we plot the relation between $\Sigma_{SFR}$ and $\Sigma_{gas}$ for the local cloud sample.  It is very clear from the figure that there is no {\it global} Schmidt law for (i.e., between) local GMCs. This is readily understood by considering the basic physical properties of galactic GMCs. In particular, the well known scaling law between cloud size and mass, $M_{GMC} = \Sigma_{A_0} R_{GMC}^2$,  first documented by Larson (1981).  Here $\Sigma_{A_0}$ is a constant which depends on the parameter $A_0$, the extinction defining the outer boundary of the cloud (Lombardi et al. 2010). For a cloud boundary starting at $A_K$ $\geq$ 0.1 magnitude,  Lombardi et al. (2010) found this constant to be  41 $\pm$ 4 \msun\ pc$^{-2}$ as can be ascertained from the figure. A similar value (42 {$\pm$} 37  \msun\ pc$^{-2}$) has been determined for a larger and more distant sample of galactic GMCs from $^{13}$CO observations by Heyer et al. (2009) who in addition found no evidence for a systematic variation in $\Sigma_{A_0} $ with galactocentric radius over a range of 4 - 8 kpc.  Galactic molecular clouds are apparently characterized by a constant gas column density that corresponds to $A_V$ $\approx$ 2 magnitudes. This fact can be understood theoretically as a result of the requirement of a minimum threshold necessary for the clouds to self-shield against molecular dissociation and  the photoionization feedback from star formation necessary to keep $\Sigma_{gas}$ from increasing too far beyond the self-shielding threshold (McKee 1989). Thus, in galactic clouds the value of $\Sigma_{SFR}$ must vary independently of  $\Sigma_{gas}$. There is no Schmidt scaling relation between molecular clouds. 

\subsection{Star Forming Galaxies}

It is of interest to place our results in the context of extragalactic studies where the Schmidt law plays an important role in investigating galaxy evolution across cosmic time. The most comprehensive measurements of the Schmidt law are those of Kennicutt (1998 and references therein, see also Kennicutt \& Evans 2012) who compiled galaxy-averaged measurements of the SFR and total gas  (i.e., atomic \& molecular) surface densities for a large sample of star forming galaxies including normal spirals and starbursts. He derived  the empirical scaling law,  $\Sigma_{SFR} \propto \Sigma_{gas}^{\rm n}$, for the galaxy-averaged $\Sigma$s, finding ${\rm n} = 1.4$.  This relation is known as the Kennicutt-Schmidt (KS) law. Of particular interest here are the resolved observations of nearby disk galaxies where the SFRs and molecular (H$_2$) gas surface densities are averaged over 1 kpc sub-regions of the galaxies and the resulting KS law has been found to have an index,  ${\rm n }= 1.0$ (Bigiel et al 2008, Schruba et al. 2011).  But how does the extragalactic KS law relate to the local Schmidt law for GMCs ($\Sigma_{SFR} = \kappa\Sigma_{gas}^\beta$)?
For the reasons outlined below we will argue that, in general, $\beta$ $\ne$  n and the index of the KS law  does not represent any underlying law of star formation, at least on spatial scales in excess of 1 kpc in star forming disk galaxies. 

In Figure 8 we plot the locus or range that is occupied by  measurements of $\Sigma_{SFR}$ and $\Sigma_{H_2}$, the molecular gas surface density, averaged over 1 kpc sized regions within a sample of nearby galaxies obtained by Schruba et al. (2011). Unlike the local GMCs, these extragalactic measurements continuously span a large range in $\Sigma_{gas}$, covering  over two orders of magnitude. Moreover, as mentioned above, there seems to be a Kennicutt-Schmidt relation for these galaxies with n $= 1$. But how can this be possible if molecular clouds are characterized by a constant mass surface density?  How is it then that molecular gas surface densities are found to range between 0.3 - 100 \msun \ pc$^{-2}$ in Schruba et al.'s observations? 

The answer to these questions can be found in the measurements at the low end of the molecular gas surface density range. Molecular gas is clearly detected in these galaxies at  surface densities as low as  0.3 - 1 \msun\ pc$^{-2}$ which corresponds to  {\it visual} extinctions of only $A_V$ $\approx$ 0.01 - 0.04 magnitudes. Since column densities of at least 1 magnitude of visual extinction are required  for dust and self-shielding to protect molecules from being dissociated and ionized by the background UV radiation field, such clouds should not exist. Clearly the CO observations used to determine the gas surface densities in these galaxies must be heavily beam-diluted and the surface densities severely underestimated. The Schruba et al. (2011) measurements are averaged over a spatial scale of 1 square kpc (10$^6$ pc$^2$) and given the typical areas of individual GMCs of 300-1000 pc$^2$, it is perhaps not surprising that significant beam dilution characterizes these measurements  (Leroy et al. 2009).  If one posits that extragalactic clouds are also characterized by a constant average column density like galactic clouds, then the linear sequence of these extragalactic star forming regions in the $\Sigma_{SFR}$-$\Sigma_{gas}$ plane can be explained as a natural consequence of measurements sampling a continuous range of beam dilutions from 0.01 to 1.0 across the galaxies. This point has been nicely demonstrated in the detailed modeling analysis of Calzetti et al. (2012).
Therefore, the measured slope of observed Kennicutt-Schmidt relation in disk galaxies is likely an artifact of unresolved observations of star forming regions and thus  not a result of any underlying physical law of star formation operating within GMCs. 

Expressing unresolved measurements of the SFRs and gas masses as surface densities necessarily introduces the dilution of the sought after physical quantities and is a general drawback for extragalactic studies of the Schmidt law. Diluted gas surface densities  of star forming regions are in essence measurements of the surface density of clouds rather than the gas within the clouds. However translating such measurements into useful information about the distribution of star forming clouds involves unravelling a complex web of factors such as the stochastic sampling of the cloud mass function, the relation between SFR and cloud mass,  intrinsic variations in $\Sigma_{A_0}$, etc. (Calzetti et al 2012). Thus, care must be exercised in interpreting the Kennicutt-Schmidt relation derived from such observations. This situation can be largely alleviated by investigating scaling relations between the integrated quantities of total SFR and gas mass, (i.e, $SFR \propto M_{gas}^p$),  such as in the studies of  Lada et al. (2010, 2012), Gao \& Solomon (2004), and Wu et al. (2005). Such measurements do not suffer from the effects of beam dilution and offer more direct insights into the physical process of star formation in regions whose clouds are unresolved by the observations being analyzed. 
Of course  resolved observations of star formation regions in nearby galaxies with ALMA will also be able to help remedy this situation. In particular, ALMA observations should be able to more clearly ascertain whether Larson's scaling relation applies to other galaxies and, if so,  the extent to which  $\Sigma_{A_0}$ can vary in differing environments.

%We note that this caveat strictly applies to only disk galaxies since these are the objects we have considered here. However, most of the galaxies in typical Kennicutt-Schmidt samples are starbursts and the physical nature of gas in these systems is likely quite different from GMCs in galactic disks. Nonetheless, observations of the gas in starbursts are also beam diluted and to some extent the concerns raised here may apply to these objects as well. 

\section{Summary and Concluding Remarks}

We have applied the Bayesian methodology recently introduced  by Lombardi et al. (2013) to investigate  the conjecture of Schmidt (1959) that the star formation rate scales with the gas density as a power-law within local molecular clouds. Our primary conclusions are as follows:

\noindent
1) We find that a local Schmidt scaling relation of the form $\Sigma_*(A_K) = \kappa A_K^\beta$ (stars pc$^{-2}$)  exists  within the four local clouds we studied. 

\noindent
2) We find very similar values of $\kappa$ and $\beta$ in three of the clouds (Orion A, Taurus, \& California) in our sample with weighted averages of $\kappa = $1.70 $\pm$ 0.01 and $\beta =$
2.04 $\pm$ 0.01, while for the fourth cloud (Orion B) we find  the significantly different values of $\kappa$ $=$ 0.77 $\pm$ 0.11 and $\beta = $ 3.30 $\pm$ 0.21. 

\noindent
3) We find that the Schmidt scaling law by itself is neither sufficient to describe, nor to accurately predict,  the level star formation activity within molecular clouds. We demonstrate that the structure of a cloud plays a crucial role in setting the level of its star forming activity. We show that the total number of protostars formed in a cloud (or equivalently, the total SFR) is proportional to the product of $\Sigma_*(A_K)$ and $S'(>A_K)$, (the differential area distribution function characterizing the cloud), integrated over all extinctions in the cloud. So for clouds with a similar functional form  for $\Sigma_*(A_K)$,  observed differences in their overall star formation activity are primarily a result of differences in $S'(>A_K)$ between the clouds.  These differences can be large in galactic molecular clouds and they  arise from a combination of differences in the overall sizes of the clouds and in the internal distribution of extinction within them.  Because of the non-linear rise in $\Sigma_*(A_K)$ with extinction, a cloud's star formation rate is particularly sensitive to the fraction of the cloud area that exists at high extinction.  

\noindent
4) The increasingly steep decline and ultimate truncation of \sagta  (the cummulative area distribution function) at high extinction, however,  effectively curtails star formation at the highest extinctions and imposes an extinction scale on an otherwise scale-free Schmidt power-law for star formation.  For relatively steep values of $\beta$ (i.e., $\geq$ 2) this results in significantly enhanced protostar production and star formation rates at extinctions in excess of \ak $\sim$ 0.8 magnitudes and explains the results of recent observational studies (Lada et al. 2010, Evans et al. 2013) that showed that the SFR in GMCs and dark clouds scales most directly with cloud mass above $\sim$ 0.8 magnitudes of extinction. The jump in SFR across this boundary can be very steep, as illustrated by the ratio of the mean protostellar  surface densites,  $\Sigma_*(\geq 0.8)/\Sigma_*(<0.8)$ $=$ 10-100, found here and in the recent study of dark clouds by Evans et al. (2013).  

\noindent
5) Two of the clouds in our sample show evidence for  Heaviside threshold extinctions in their internal Schmidt relations at $A_0$ $\approx$ 0.25 and 0.60 magnitudes. The two other sources show no evidence for such thresholds consistent with earlier studies (e.g., Gutermuth et al. 2011).   However, we find that the presence or absence of such thresholds in the \sfrsd\ of the Schmidt law has little to do with the steep jump in protostellar production and (integrated) SFR observed across the 0.8 magnitude extinction boundary. 

\noindent
6) None of the clouds in our study showed any evidence for detectable diffusion of protostars from their birth sites.

\noindent
7) We demonstrate that there is no Schmidt scaling law describing star formation between clouds and argue that this is a natural consequence of the well-known scaling law between mass and size of molecular clouds that was first described by Larson (1981).  

\noindent
8) Unresolved (i.e., S $>$ 1 kpc$^2$) measurements of  $\Sigma_{gas}$ in disk galaxies primarily measure the surface densities of clouds rather than the gas; thus the observed functional relation between $\Sigma_{SFR}$ and $\Sigma_{gas}$  (i.e., the Kennicutt-Schmidt law) is not the result of any underlying physical law of star formation operating within molecular clouds.

In summary, our analysis of the star formation scaling relations for four nearby molecular clouds  demonstrates that Schmidt's original conjecture applies to star formation within but not between local molecular clouds when the scaling relation is expressed in the areal form that relates the protostellar and gas surface densities. It is interesting that even though individual GMCs can be characterized by internal Schmidt-like scaling relations for star formation, these relations do not provide a complete predictive description of the star formation activity in such clouds. Such a description does become possible once the effect of the structure of the cloud is coupled to the Schmidt scaling law. This results in a modified scaling law between the cumulative protostellar fraction or SFR and cloud surface density. This modified star formation law can also account for the observed correlation of the total SFR with the mass of high extinction (or dense) material in Galactic GMCs and provides a framework for a potentially  deeper understanding of extragalactic observations of star formation.

Finally, one may question whether the internal Schmidt scaling relation between the SFR and  mass surface densities  represents the most physically meaningful star formation law for a cloud. For one thing, the measured surface densities represent  two dimensional projections of both the SFR and cloud mass and thus should  depend on the cloud orientation to the line-of-sight, and this is not a desirable property for a general physical law of star formation. Indeed, any observed variations in the measured parameters of the Schmidt relations within Galactic clouds could be a result of such geometrical factors. Furthermore, the theoretical underpinnings of such a law are unclear.  Existing theories of star formation are predicated on the basic idea that star formation results from an imbalance between the inward pull of gravity and the outward push of internal pressure within a molecular core or cloud, as expressed by the Jeans' inequality   ( i.e., $M_{\mathrm gas} > M_J \approx {4\pi c_s^3 \  /  \ 3(G^3\rho)^{1\over 2}} $ for gas with sound speed, $c_s$).  
Therefore, on theoretical grounds, a volumetric Schmidt law (e.g., $\rho_{\mathrm SFR}= \kappa\rho_{\mathrm gas}^\alpha$) would likely  be more directly related to the physical process of star formation within a cloud than the standard areal version of the relation studied here. Various theoretical studies have proposed that with $\alpha \approx 1.6$, a volumetric scaling relation could account for the observed Kennicutt-Schmidt relation for galaxies (e.g., Elmegreen 2002; Krumholz, Dekel \& McKee 2011). Furthermore, Krumholz et al. (2011) suggested such a law could simultaneously explain the local as well as extragalactic Schmidt scaling relations provided that, independent of the star formation environment, the fraction of gas going into stars within the corresponding free fall time was a constant with a value of about 1-2 \%, as was suggested in earlier theoretical work by Krumholz \& McKee (2005). However given the caveat (8) expressed above, it is not clear that comparing such theoretical predictions against the measured, beam-diluted, extragalactic surface densities constitutes a valid test of such models. If a volumetric Schmidt law exists, then it may be best to infer its parameters empirically. This could be accomplished with more detailed studies of local molecular clouds if such studies are able to provide knowledge of both the cloud-to-cloud variations in the local (areal) Schmidt law and the geometries and orientations of the clouds relative to earth.  Nonetheless, our results suggest that even a volumetric Schmidt relation, by itself, would not provide an adequate description of star formation in a cloud. It would need to be coupled to a corresponding volume density distribution function  to fully and accurately describe star formation within the cloud.

\acknowledgments
We thank Bruce Elmegreen, Neal Evans, and Chris McKee for carefully reading an earlier draft of this paper and providing comments and suggestions which resulted in improvements to this final version. We acknowledge comments from an anonymous referee which led to refinements in our arguments and improved the overall presentation of the paper. We are grateful to Mark Reid, and Leo Blitz for informative discussions. One of us (CRZ) gratefully acknowledges support from the following programs: CONACYT 152160 and DGAPA-PAPIIT-UNAM IA101812 Mexico.

\clearpage

\begin{table}
\begin{center}
\caption{Derived Scaling Law Parameters\tablenotemark{1} \label{tbl-1}}
\vskip 0.1in
\begin{tabular}{lcccc}
\tableline\tableline
Parameter & \multicolumn{1}{c}{Orion A\tablenotemark{a}} & \multicolumn{1}{c}{California} &\multicolumn{1}{c}{ Taurus}& \multicolumn{1}{c}{Orion B}  \\
\tableline
$\beta$ &2.03\  $\pm$ 0.08 &1.99\  $\pm$ 0.32  &2.09  $\pm$ 0.14 &3.30 $\pm$ 0.21 \\
$\kappa$ (stars pc$^{-2}$ mag$^{-\beta}$) &1.64\ $\pm$ 0.09 &2.05\ $\pm$ 0.40 &2.08\ $\pm$ 0.30 &0.77 $\pm$ 0.11 \\ 
$\sigma$ (pc)&0.03\  $\pm$ 0.02  &0.05\ $\pm$ 0.02  &0.01 $\pm$ 0.01 &0.04 $\pm$ 0.02 \\ 
A$_0$ (magnitudes)&0.04\ $\pm$ 0.03 & 0.62\ $\pm$ 0.04& 0.03\ $\pm$ 0.02 &0.26 $\pm$ 0.14 \\
\tableline
\end{tabular}
%% Any table notes must follow the \end{tabular} command.
\tablenotetext{1}{Errors quoted correspond to 1-$\sigma$ errors.}
\tablenotetext{a}{values for Orion from Lombardi et al. 2013}

\end{center}
\end{table}

\begin{table}
\begin{center}
\caption{Predicted and Observed Protostellar Population \label{tbl-2}}
\vskip 0.1in
\begin{tabular}{lcccc}
\tableline\tableline
%Parameter & \multicolumn{1}{c}{Orion A\tablenotemark{a}} & \multicolumn{1}{c}{California} &\multicolumn{1}{c}{ Taurus}& \multicolumn{1}{c}{Orion B  \\
Population & \multicolumn{1}{c}{Orion A} & \multicolumn{1}{c}{California} & \multicolumn{1}{c}{ Taurus} & \multicolumn{1}{c}{Orion B} \\
\tableline
Observed & 329 & 54 &51 &90\\
Predicted\tablenotemark{a} &332.7 & 55.4 & 52.1&90.5 \\
\tableline
\end{tabular}
%% Any table notes must follow the \end{tabular} command.
\tablenotetext{a}{$N_*= \int \Sigma_*(A_\mathrm{K}) \, \diff S \ {\rm and}\  \Sigma_*(A_K) = A_0 + \kappa A_K^\beta$}

\end{center}
\end{table}

\clearpage

%% the file, uncomment the \plotone, \plottwo, and
%% If you need a layout that cannot be achieved with \plotone or
%% \plottwo, you can invoke the graphicx package directly with the
%% \includegraphics command or use \plotfiddle. For more information,
%% please see the tutorial on "Using Electronic Art with AASTeX" in the
%% documentation section at the AASTeX Web site,
%% http://www.journals.uchicago.edu/AAS/AASTeX.
%%
%% The examples below also include sample markup for submission of
%% supplemental electronic materials. As always, be sure to check
%% the instructions to authors for the journal you are submitting to
%% for specific submissions guidelines as they vary from

%% This example uses \plotone to include an EPS file scaled to
%% 80% of its natural size with \epsscale. Its caption
%% has been written to indicate that additional figure parts will be
%% available in the electronic journal.

\clearpage

\begin{figure}
\includegraphics[width=\linewidth]{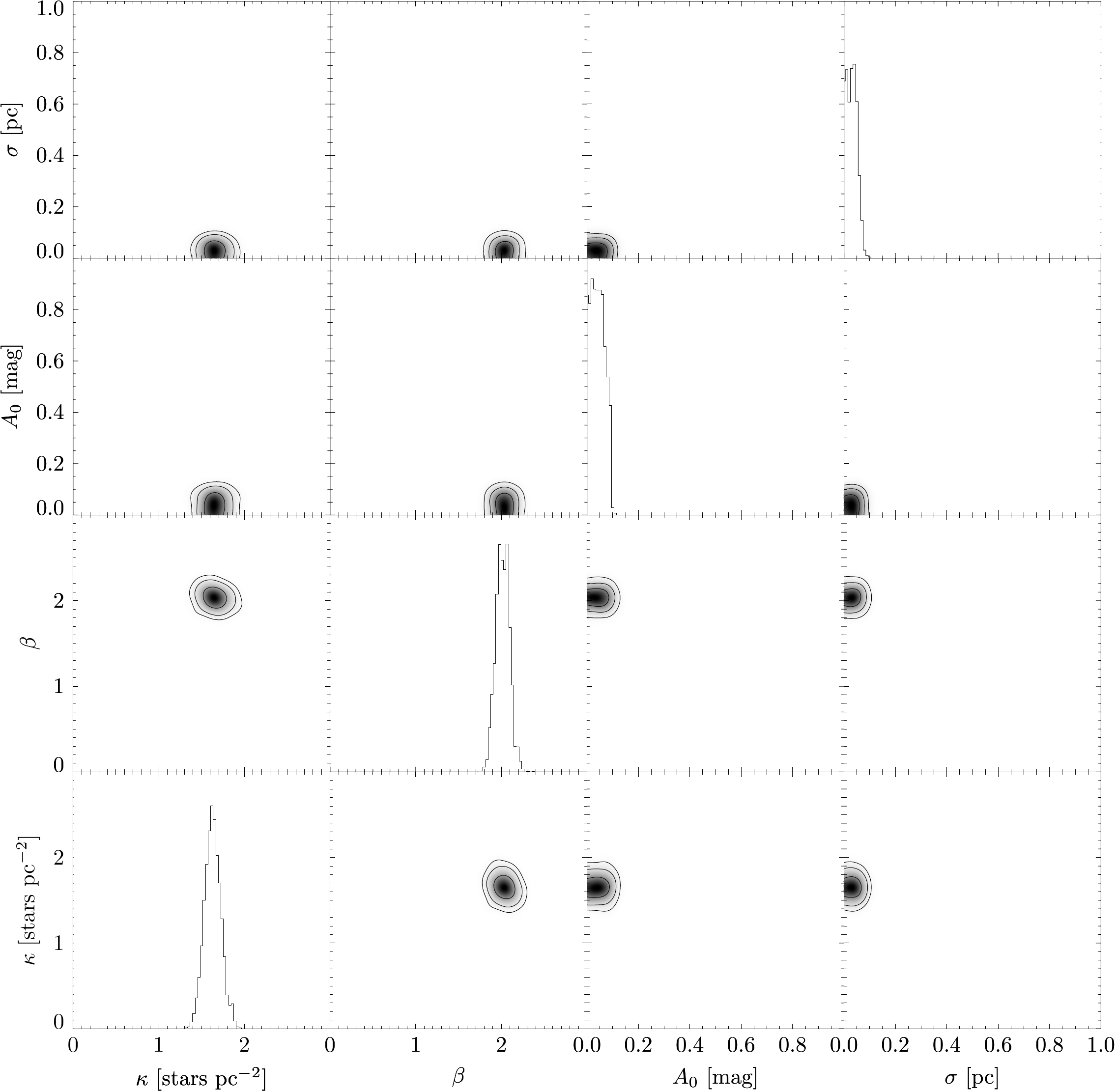}
\caption{The posterior probabilities for all combinations of the four model parameters for the protostellar population in the Orion A cloud. The contours mark the  99.7\% , 95.5\% and 68.3\% confidence levels, respectively. The boxes along the diagonal display the individual frequency distributions of probabilities for the corresponding parameters and are arbitrarily normalized on the vertical axis. The values for the parameters in the Orion A cloud are the most tightly constrained of the clouds we studied because the relatively large protostellar population in that cloud provides robust statistics.  \label{fig1}}
\end{figure}

\begin{figure}
\includegraphics[width=\linewidth]{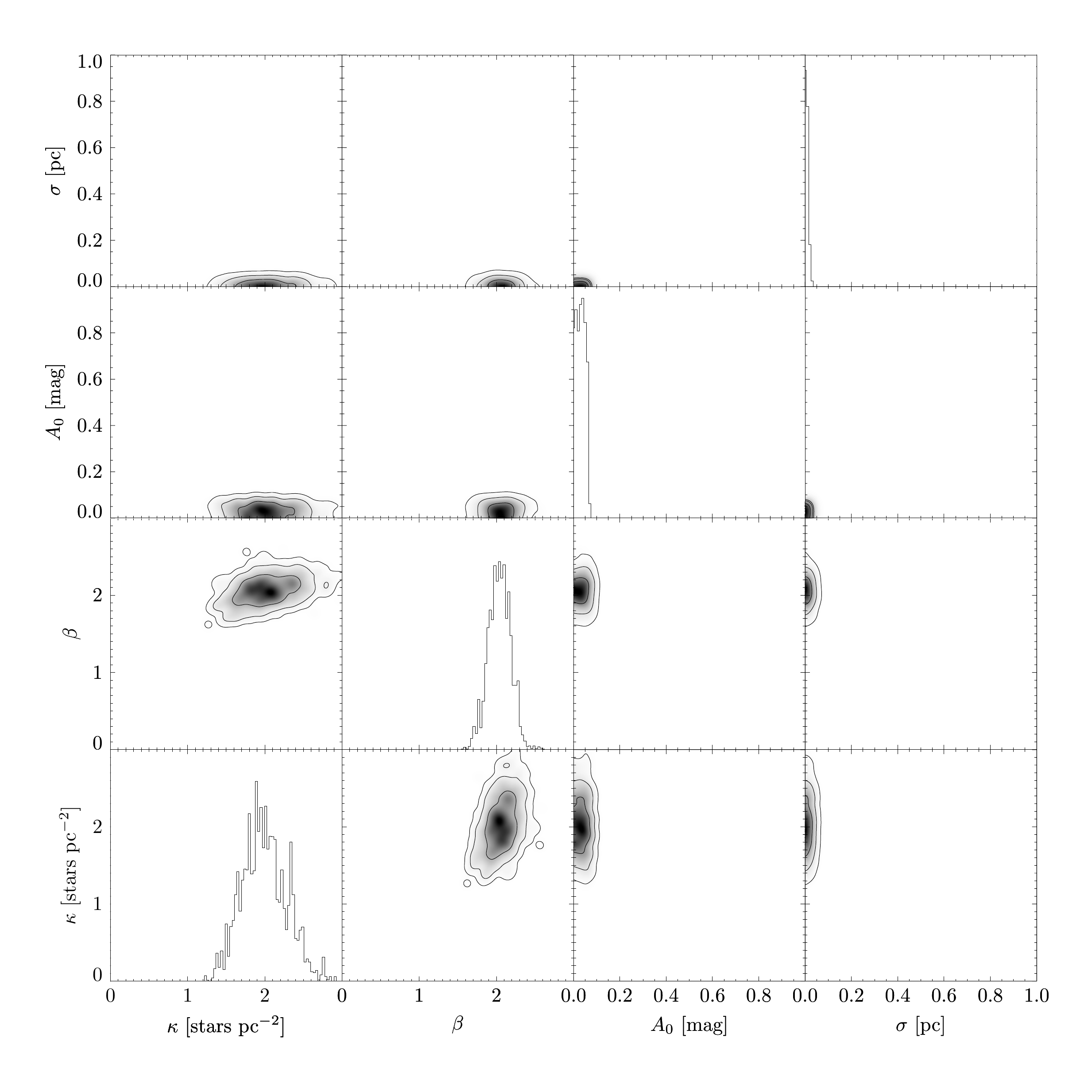}
\caption{The posterior probabilities for all combinations of the four model parameters for the protostellar population in the Taurus cloud. Otherwise same as Figure 1.  The most probable values of the parameters $\beta$, $\gamma$, $\sigma$ and $A_0$ are essentially the same as seen in the Orion cloud in Figure 1 although somewhat less tightly constrained due to the smaller protostellar population in Taurus.\label{fig1}}
\end{figure}

\begin{figure}
\includegraphics[width=\linewidth]{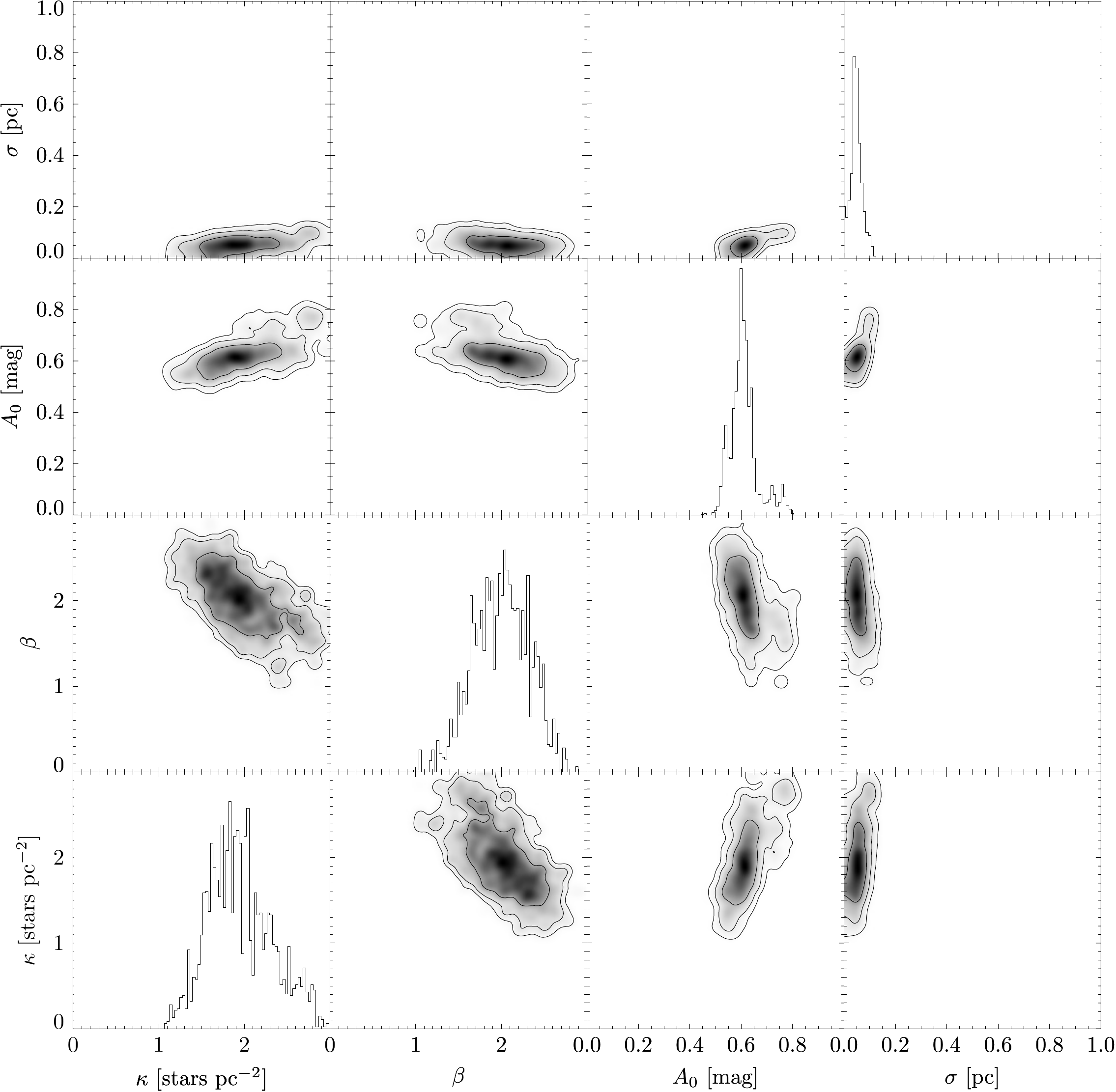}
\caption{The posterior probabilities for all combinations of the four model parameters for the protostellar population in the California cloud. Otherwise the same as Figure 1.  The most probable values for the parameters are very similar to those for Orion A and Taurus except for $A_0$ which is significantly greater than 0  in this cloud. The parameters in the California cloud are not as well constrained as those for Orion A due to its smaller protostellar population. \label{fig1}}
\end{figure}

\begin{figure}
\includegraphics[width=\linewidth]{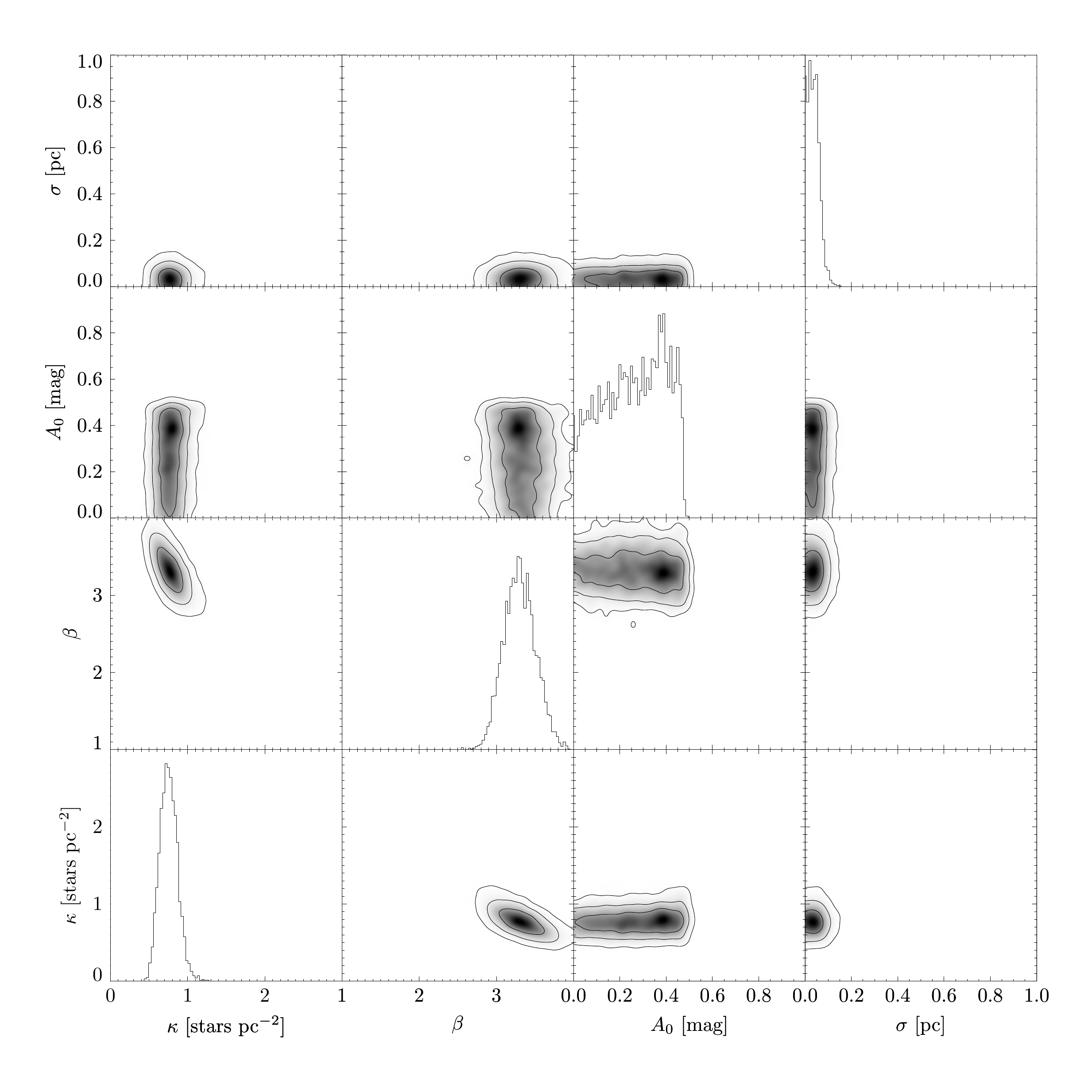}
\caption{The posterior probabilities for all combinations of the four model parameters for the protostellar population in the Orion B cloud. Otherwise the same as Figure 1.  The most probable values for the parameters $\beta$ and $\kappa$, differ significantly from those for the other three clouds. Although the power-law index $\beta$ is steeper than that found in the other clouds, the star formation coefficient, $\kappa$ is smaller. Similar to the California cloud,  there appears to be a detection of a threshold extinction, $A_0$, although with lower confidence than that derived the California cloud. The parameters in the Orion B cloud are not as well constrained as those for Orion A due to its smaller protostellar population. \label{fig1}}
\end{figure}

\begin{figure}
%\begin{center}
\includegraphics[width=0.49\hsize]{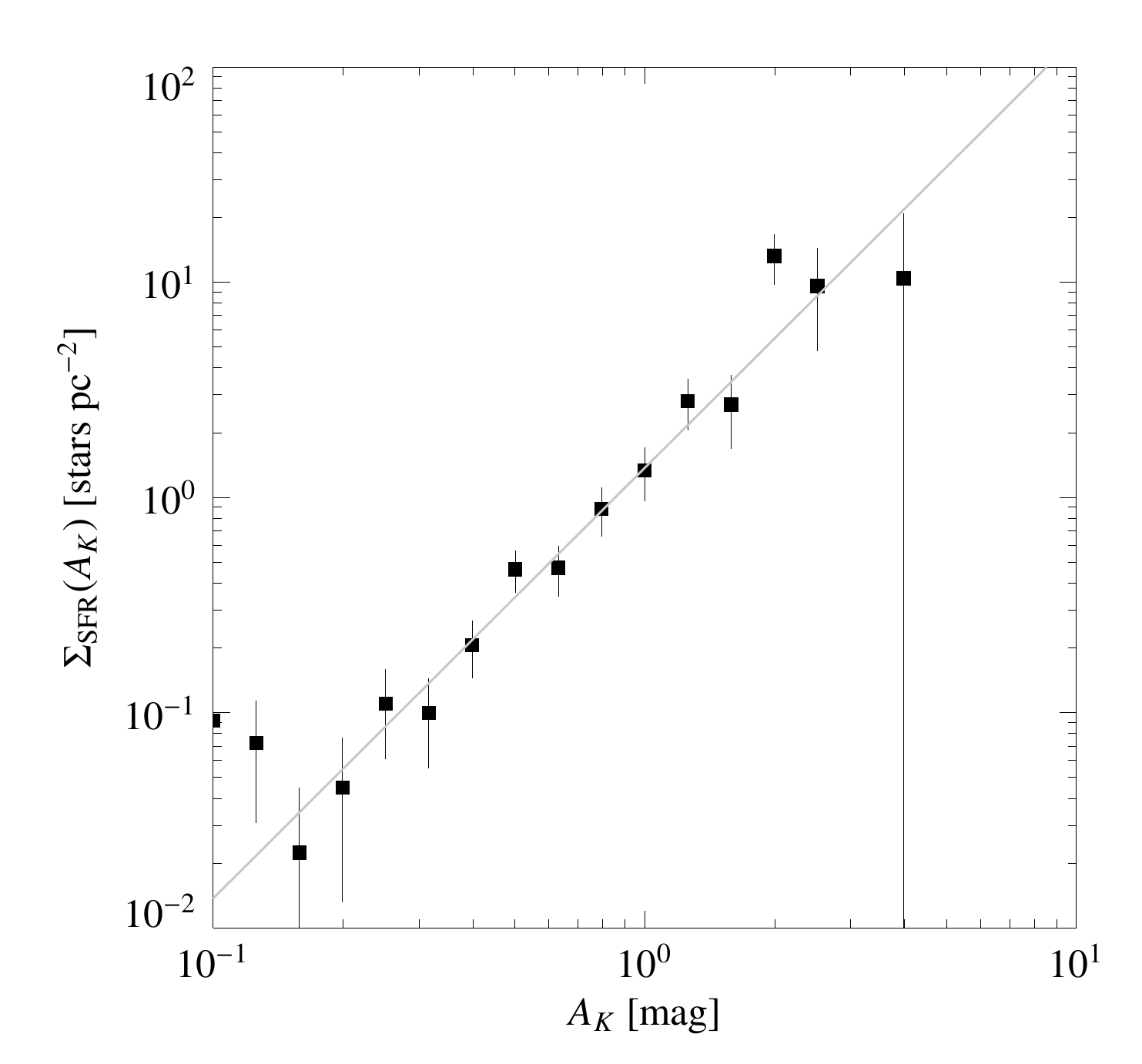}
\hfill
\includegraphics[width=0.49\hsize]{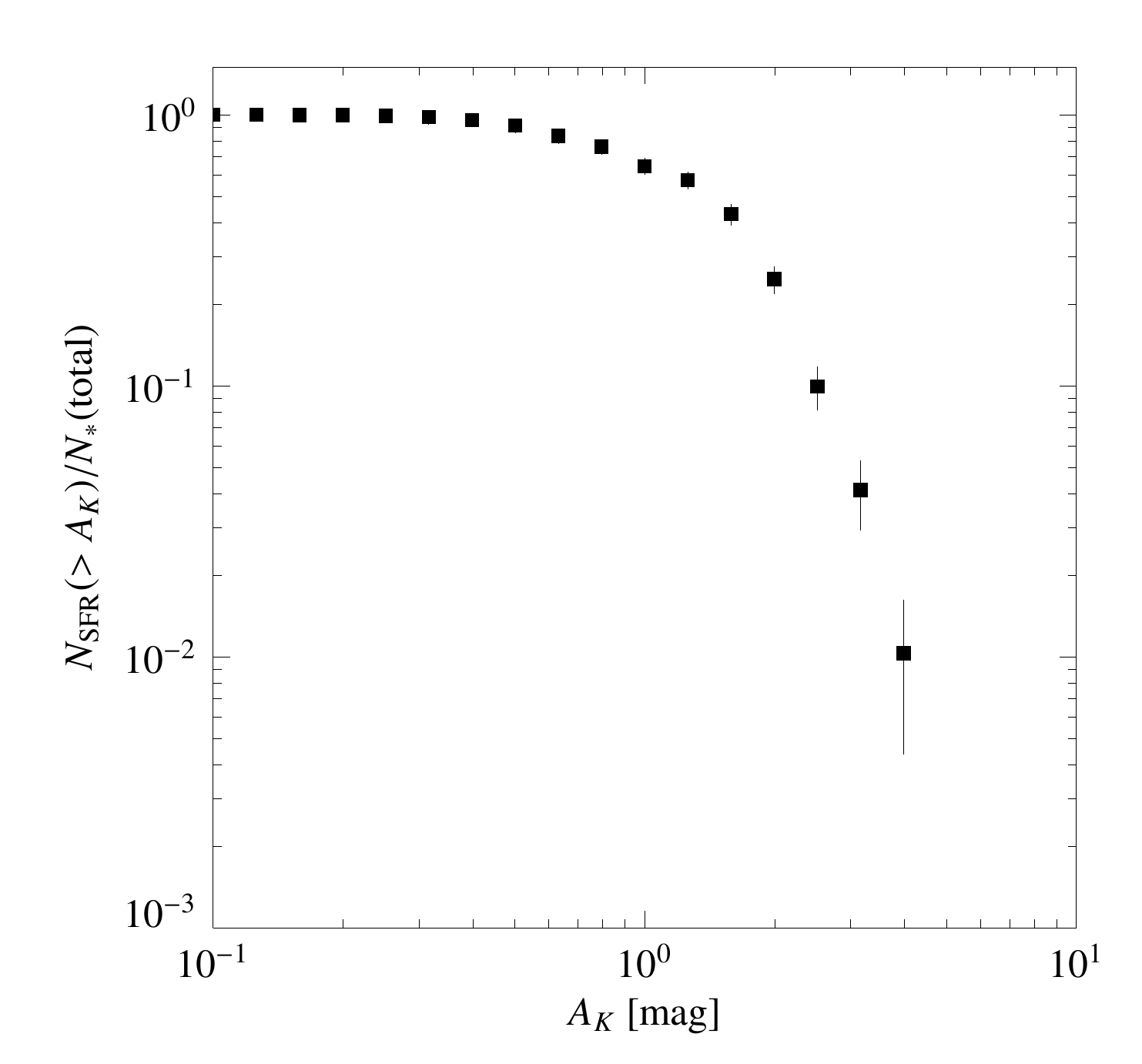}
\vskip -.2in
\caption{The left panel shows the local Schmidt scaling relation in the Orion A molecular cloud. The protostellar surface density, \psd,   steeply rises with increasing extinction in this cloud. The solid line is a least squares fit to the data which yielded a power-law index, $\beta$, of 2.0 in agreement with the Bayesian analysis (see text).  The right panel shows the variation in the cumulative protostellar fraction (CPF)  with extinction in Orion A. The function is relatively flat in the lower extinction regions that make up the bulk of the cloud. However, it drops steeply at extinctions in excess of 1 magnitude in spite of the apparently unabated, non-linear rise in $\Sigma_*$  with extinction. This behavior in the CPF is a consequence of the steep fall off of cloud area with extinction seen in Figure 6. Nonetheless, 80\% of the protostars in Orion A are found at extinctions in excess of 0.8 magnitudes. \label{fig5}}
%\end{center}
\end{figure}

\begin{figure}
\includegraphics[width=\linewidth]{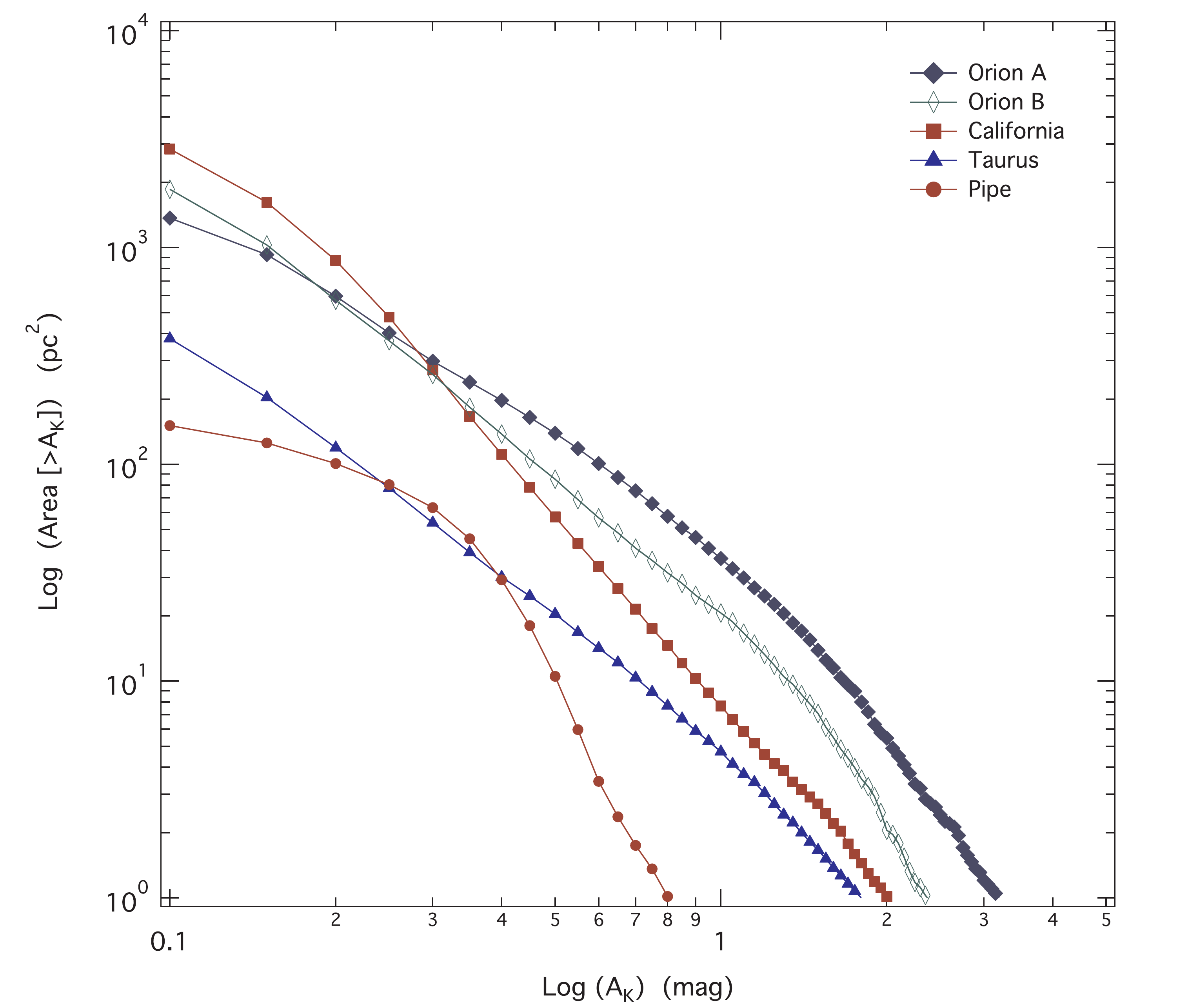}
\caption{The dependence of the area distribution function $S(>A_K)$ on column density, $A_K$, for the three clouds in this study and the Pipe Nebula for comparison. The functions, $S(>A_K)$,  all differ in amplitudes and are all decreasing functions of extinction that fall most steeply at the highest extinctions. \label{fig5}}
\end{figure}

\begin{figure}
\includegraphics[width=0.85\hsize]{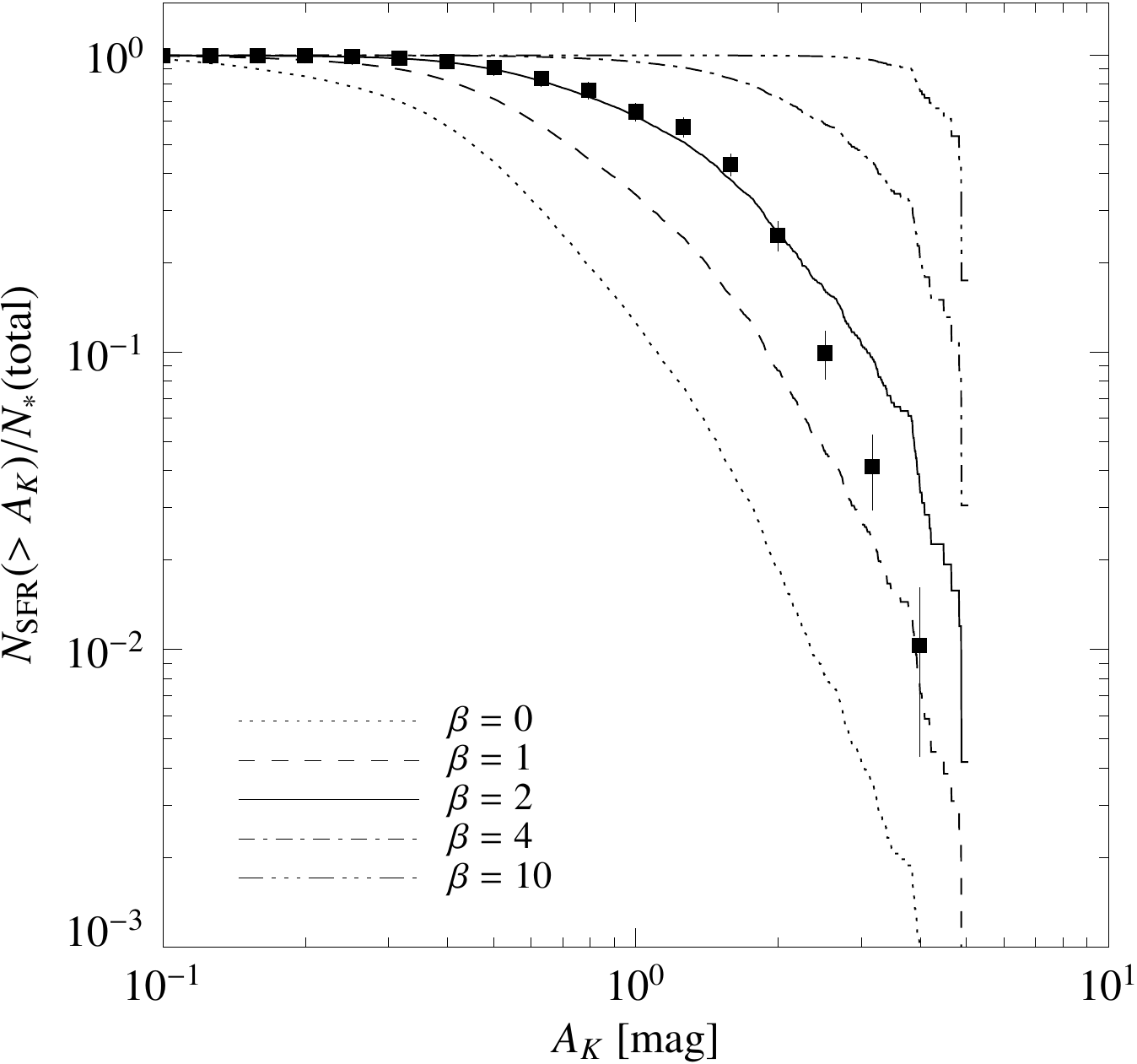}
\caption{The predicted fractional yield of protostars (or fractional SFR) as a function of infrared extinction (continuous curves). These cumulative protostellar fractions were derived from Equation 10 for a set of internal Schmidt laws with varying spectral indices $\beta$ ($=$ 0, 1, 2, 4, and 10, respectively) and assuming the area distribution function, \sagtans, measured for the Orion A molecular cloud. The bottom most curve represents the $\beta = 0.0$ case and corresponds to a cloud with a constant $\Sigma_*$. It is identical in shape to the area distribution function for Orion (e.g., Figure 6). The uppermost curve corresponds to the case of an extreme Schmidt law with an extremely  steep rise to high extinctions that resembles a sharp threshold for the fractional the star formation rate. Here the threshold appears to be just above 4 magnitudes of extinction (see text). Also plotted are the observations of the Orion A cloud for comparison. The Orion A observations follow the model prediction ($\beta$ $=$ 2) fairly closely below a dust column density of \ak $\approx$  2.0 magnitudes (i.e., $\Sigma_{gas}$ $\approx$ 393 \msun\  pc$^{-2}$). At higher extinctions the data fall below the predicted relation likely due to an observational bias resulting from the small number statistics that characterize the last three bins and the resulting correlated nature of the source counts in these bins. \label{fig7a}}
\end{figure}

\begin{figure}
\includegraphics[width=\linewidth]{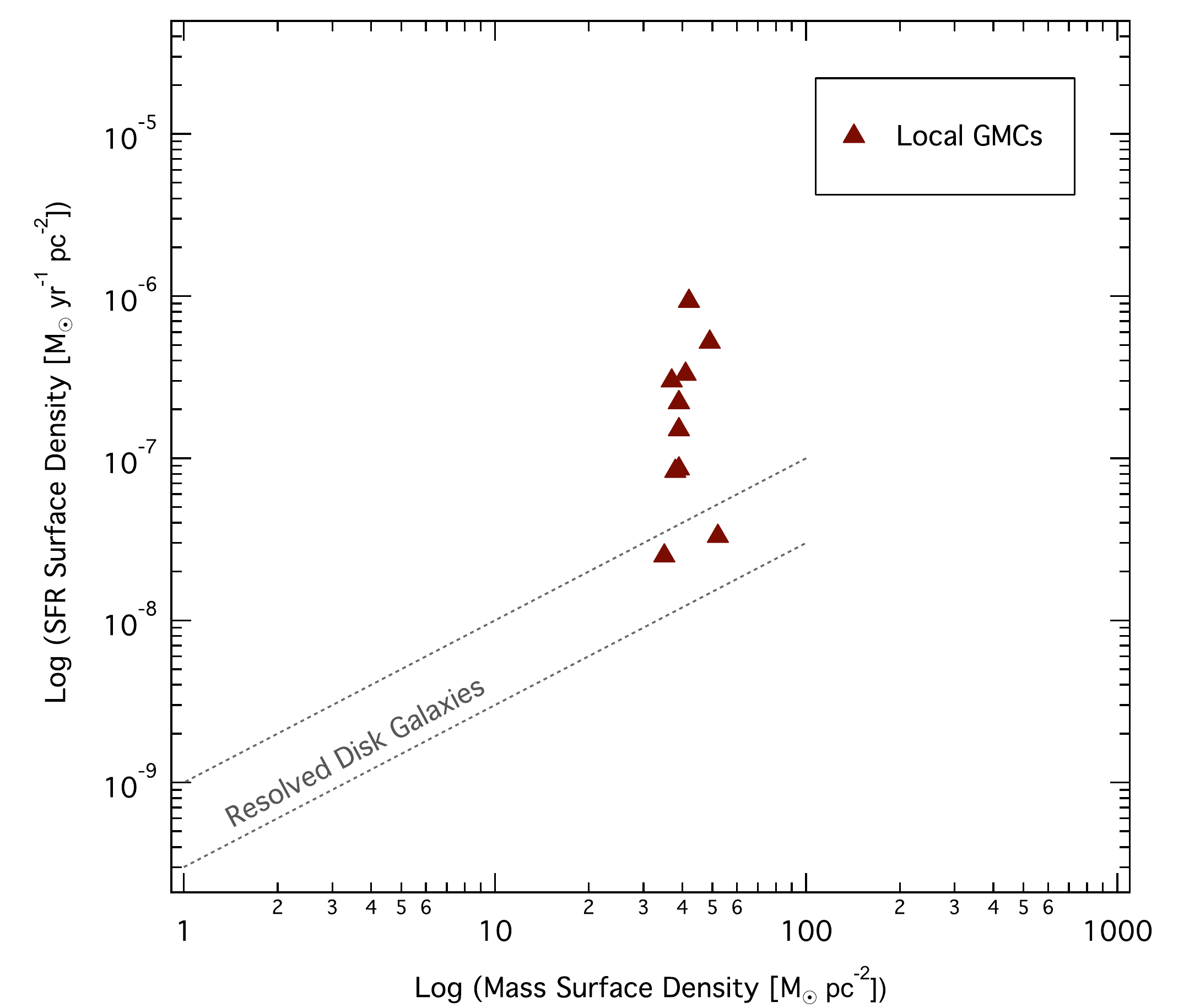}
\caption{The $\Sigma_{SFR}$ vs $\Sigma_{gas}$ relation for local GMCs. The plot shows that a Kennicutt-Schmidt law, i.e., $\Sigma_{SFR} \propto \Sigma_{gas}^n$, does not exist for local molecular clouds.  Also plotted is the regime occupied by resolved measurements of nearby disk galaxies (Schruba et al. 2009) which do show a Kennicutt-Schmidt relation with $n = 1$. (see text) \label{fig6}}
\end{figure}

%% If you are not including electonic art with your submission, you may
%% mark up your captions using the \figcaption command. See the
%% No more than seven \figcaption commands are allowed per page,
%% so if you have more than seven captions, insert a \clearpage

%% Tables should be submitted one per page, so put a \clearpage before

%% Two options are available to the author for producing tables:  the
%% deluxetable environment provided by the AASTeX package or the LaTeX
%% table environment.  Use of deluxetable is preferred.
%%

%% Three table samples follow, two marked up in the deluxetable environment,
%% one marked up as a LaTeX table.

\end{document}